\documentclass[aps,prb,twocolumn,showpacs,superscriptaddress,floatfix]{revtex4-1}

\pdfoutput=1
\usepackage[titletoc,toc,title]{appendix}
\usepackage{amsmath,amssymb}
\usepackage{graphicx}
\usepackage{color}
\usepackage{rotating}
\usepackage{bm}
\usepackage{setspace}
\usepackage{hyperref}
\usepackage{float}
\usepackage{soul}
\usepackage[T1]{fontenc}
\usepackage{bbold}
\usepackage{color}
\usepackage{braket}

\hypersetup{
colorlinks=true,
urlcolor= blue,
citecolor=blue,
linkcolor= blue,
bookmarks=true,
bookmarksopen=false,
}

\renewcommand{\vec}[1]{\bm{#1}}

\begin{document}

\preprint{APS/123-QED}

\title{Angle-Dependent {\it Ab initio} Low-Energy Hamiltonians for a Relaxed Twisted Bilayer Graphene Heterostructure}
\author{Shiang Fang}
\affiliation{Department of Physics, Harvard University, Cambridge, Massachusetts 02138, USA}
\affiliation{John A. Paulson School of Engineering and Applied Sciences, Harvard University, Cambridge, Massachusetts 02138, USA}
\author{Stephen Carr}
\affiliation{Department of Physics, Harvard University, Cambridge, Massachusetts 02138, USA}
\author{Ziyan Zhu}
\affiliation{Department of Physics, Harvard University, Cambridge, Massachusetts 02138, USA}
\author{Daniel Massatt}
\affiliation{Department of Statistics, The University of Chicago, Chicago, Illinois 60637, USA}
\author{Efthimios Kaxiras}
\affiliation{Department of Physics, Harvard University, Cambridge, Massachusetts 02138, USA}
\affiliation{John A. Paulson School of Engineering and Applied Sciences, Harvard University, Cambridge, Massachusetts 02138, USA}

\begin{abstract}
We present efficient angle-dependent low-energy Hamiltonians to describe the properties of 
the twisted bilayer graphene (tBLG) heterostructure, based on {\it ab initio} calculations of mechanical relaxation and electronic structure. 
The angle-dependent relaxed atomic geometry is determined by continuum elasticity theory, which induces both in-plane and out-of-plane deformations in the stacked graphene layers. 
The electronic properties corresponding to the deformed geometry are 
derived from a Wannier transformation to local interactions obtained from 
Density Functional Theory calculations. 
With these {\it ab initio} tight-binding Hamiltonians of the relaxed heterostructure, 
the low-energy effective theories are derived from the projections near 
Dirac cones at K valleys. 
For twist angles ranging from 0.7$^\circ$ to 4$^\circ$, we extract
both the intra-layer pseudo-gauge fields and the inter-layer coupling terms 
in the low-energy Hamiltonians, 
which extend the conventional low-energy continuum models. 
We further include the momentum dependent inter-layer scattering terms 
which give rise to the particle-hole asymmetric features of the electronic structure. 
Our model Hamiltonians can serve as a starting point for formulating 
physically meaningful, accurate interacting electron theories.
\end{abstract}

\maketitle

\section{Introduction}

The physical system consisting of two layers of graphene with a small 
relative twist between them, referred to as 
twisted bilayer graphene (tBLG), has emerged as a new platform 
for studying correlated phases of matter since the discovery 
of its Mott insulator~\cite{TWBLG_Mott} and the superconducting~\cite{TWBLG_SC} behavior. 
The unconventional nature of both the insulating and superconducting phases 
has prompted further experimental~\cite{TWBLG_pressure_EXP,TWBLG_strange_metal,TWBLG_STM1,TWBLG_STM2,TWBLG_FM_EXP,TWBLG_moreSC,TWBLG_more_magic} and theoretical efforts~\cite{TwBLG_phSC1,TwBLG_phSC2,TwBLG_th1,TwBLG_th2,TwBLG_th3,TwBLG_th4,TwBLG_th5,TwBLG_th6,TwBLG_th7,TwBLG_th8,TwBLG_th9,TwBLG_th10,TwBLG_th11} 
to better understand the physics of tBLG and related 
van der Waals heterostructures. 
Unconventional correlated phases are also observed in the heterostructures
that include trilayer graphene with nearly aligned hBN substrate~\cite{triGhBN_Mott,triGhBN_SC} and twisted double bilayer graphene~\cite{biGbiG1,biGbiG2,biGbiG3}. 
The hypothesis is that a better understanding
of the correlated many-body phases will emerge by exploring the many parameters 
available to tune the behavior of the stacked-layer hetersotstructures; 
these parameters include doping, external electric and magnetic fields, 
temperature and applied pressure~\cite{TWBLG_pressure_EXP}. 
The response of the system to changes in the parameters 
would serve to set constraints on theoretical models, much as was the case for the isotope effect~\cite{SC_isotope}
or the effect of external magnetic fields in understanding conventional superconductivity. 
In contrast to the conventional three-dimensional crystalline solids, one unique adjustable  
parameter for tBLG and its relatives, 
is the twist angle between layers: 
by manipulating the corresponding moir\'e length scale through the twist angle, 
which can be controlled to exquisite precision, 
the characteristic kinetic and interaction energies can be varied 
without sacrificing the material quality, 
an effect referred to as ``twistronics''~\cite{Twistronics_DFT}. 
As a consequence, the existence of the unconventional, correlated phases 
depends sensitively on twist angle variations~\cite{TWBLG_Mott,TWBLG_SC,TWBLG_pressure_EXP}. 
The flat bands that emerge in the electronic spectrum at the ``magic angle''
($\sim 1^{\circ}$ for tBLG)
are a characteristic feature of these systems that signals the emergence of 
strong electron interaction effects, as first pointed out by 
A. McDonald and coworkers~\cite{MacDonald_KP}. 

At the single-particle theory level, accurate theoretical modeling 
can capture the angle-dependent effects on the electronic structure and band topology~\cite{Ashvin_10band,Bernevig_TWBLG,TwBLG_top1}, 
and thus serves as a starting point for formulating minimal models of interacting theories~\cite{Ashvin_Mott,TWBLG_wan1,TWBLG_wan2,Liang_2018}.
Such theoretical modeling should take into account both the mechanical relaxation and 
the electronic properties. 
In single-layer graphene, the structural deformation can affect the 
electronic structure by inducing pseudo-gauge fields coupled to the Dirac electrons~\cite{eff_graphene_strain,TMDC_strainTBH,curve_space2,rev_graphene_gauge}. 
In the case of tBLG, the interaction between the two misaligned layers gives rise to a
modulated structural pattern~\cite{Koshino_relax,Carr_GSFE}. 
The origin of the modulation is due to the varying local geometric configurations. 
Among these configurations, the Bernal stacking order as in the bulk graphite structure 
is most favorable energetically~\cite{GSFE_1,GSFE_2,Carr_GSFE}. 
The relaxed tBLG crystal is hence determined by optimizing the stacking energy cost 
and elastic energy from layer deformation. 
Near the magic angle, the crystal relaxation can affect the electronic structure around the flat bands~\cite{Koshino_relax}. The single-particle gaps on both electron and hole sides of the flat bands can be accounted for by including crystal relaxation when compared with the experiments. When the twist angle is varied, the strength of atomic relaxation can be modified as well~\cite{Koshino_relax,Carr_GSFE}. It is therefore important to construct electronic models that capture these angle-dependent effects.


Given the relaxed crystal structure of a tBLG, one can model the electronic properties. A numerical scheme to model such van der Waals heterostructures has to be accurate enough to make numerical predictions that can be compared to experiment. 
Moreover, it should also give an intuitive physical picture that can facilitate 
the design of structures to enable electron correlations. 
The large number of atoms involved in the supercells of 
twisted bilayers has prevented straightforward numerical approaches. 
Numerical methods at different levels of sophistications from large scale density functional theory (DFT) calculations~\cite{graphene_sc_dft1,graphene_sc_dft2}, empirical tight-binding Hamiltonians~\cite{graphene_sc_tbh1,graphene_sc_tbh2,graphene_sc_tbh3,graphene_sc_tbh4,Koshino_relax} to low-energy $\vec{k} \cdot \vec{p}$ continuum theories~\cite{MacDonald_KP,graphene_sc_kp1,graphene_sc_kp2,graphene_sc_kp3,graphene_sc_kp4}  have been employed. Among these approaches, the continuum $\vec{k} \cdot \vec{p}$ is computationally efficient and allows continuous twist angle control of the electronic band structure unconstrained by commensurate conditions~\cite{Zou_TWBLG,graphene_sc_kp4}; 
the latter type of constraints
are required for DFT and tight-binding calculations. 
Each approach has its strengths and weaknesses in terms of numerical accuracy and efficiency.

In our work, we adopt an {\it ab initio} multi-scale numerical approach to model a tBLG 
which takes fully into account crystal relaxation and the twist angle dependence. 
Briefly, 
this multi-scale numerical method is designed to combine the strengths of conventional approaches mentioned above. 
This framework allows us to extract the relevant mechanical and electronic properties at the microscopic length scale based on {\it ab initio}
total-energy and electronic structure calculations. 
We thus obtain a computation scheme that allows for both 
a clear physical picture and efficient numerical implementation, 
discussed in detail in the main body of the paper. 
Through this approach, we are able to
generalize the conventional continuum  $\vec{k} \cdot \vec{p}$ theory~\cite{MacDonald_KP} 
to capture all relevant band 
features of {\it ab initio} methods~\cite{Carr_10band}. 
For example, one such feature is the pronounced particle-hole asymmetry in the tight-binding bands in tBLG, a theoretical prediction that remains an
open question on the experiment side.

Another interesting aspect of the flat bands near the Fermi level 
is the topological properties of the manifold,
related to the wavefunction texture in the full Brillouin Zone. 
In contrast with trivial atomic insulator bands which can be viewed as the Hilbert space from spatially localized orbitals, 
a non-trivial topological structure of the flat bands in a tBLG near the magic angle has implications on the construction of a finite $n$-band model near charge-neutrality~\cite{Ashvin_10band,Bernevig_TWBLG}. 
In other heterostructures, flat bands with non-trivial Chern numbers are predicted~\cite{Moire_Chern,Moire_topological}. 
The non-trivial topology in the manifold impacts the formation of exotic interacting phases. 
The accurate multi-scale modeling can provide a reliable numerical method to estimate the model parameters and investigate various perturbations and their effects on the topology of the flat bands.

The paper is structured as follows: In Sec. II, we describe the procedures in each step of the multi-scale approach, from mechanical and electronic calculations with DFT and Wannier constructions, to {\it ab-initio} tight-binding models and the derivations of effective low-energy $\vec{k} \cdot \vec{p}$ Hamiltonians based on the projection method. In Sec III, we further explore the numerical implications of our effective models which include the mechanical and electronic properties, and the scaling of coupling constants with respect to the twist angle. We conclude and summarize the discussions in Sec IV, and remark on potential applications and extensions.

\section{Relaxed tBLG Mechanical and Electronic Structure: A Multi-scale Approach}

\begin{figure}[h]
  \centering
  \includegraphics[width=.5\textwidth]{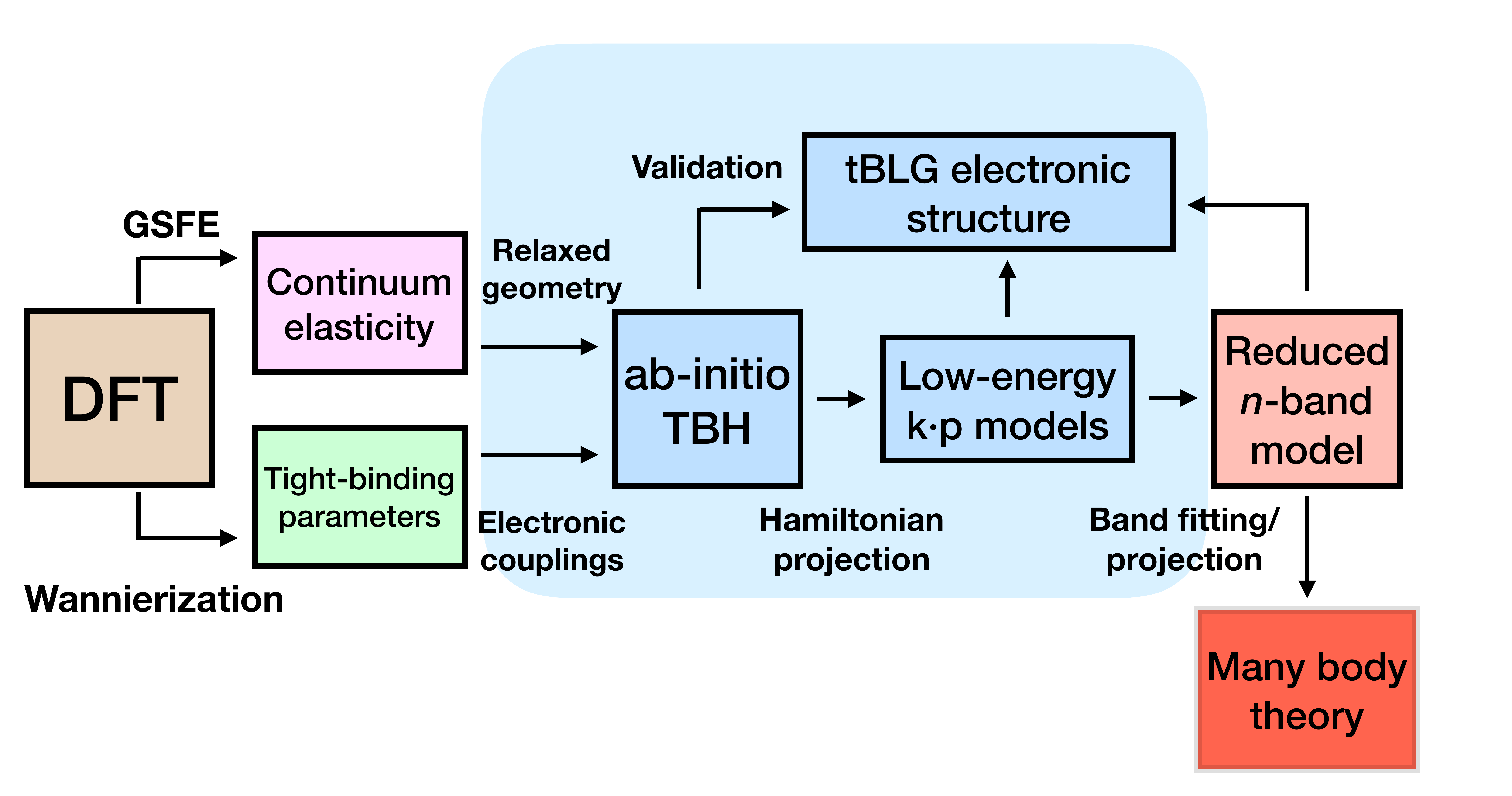}
  \caption{
Schematic representation of the multi-scale numerical approach 
for the tBLG electronic structure, which begins at the microscopic scale where 
the mechanical and electronic properties of the local configuration are 
modeled with DFT calculations. 
These results are employed for the construction of
{\it ab initio} tight-binding Hamiltonian and 
low-energy effective $\vec{k} \cdot \vec{p}$ theories. 
The reduced $n$-band models are further simplifications 
intended to form the basis for interacting many-body theories.
The blocks in blue color are the focus of the present work (included in the 
light-blue shaded region).}
  \label{TwBLG_multiscale}
\end{figure}

A van der Waals heterostructure as exemplified in a tBLG is a complicated material system to study, with massive number of atoms involved at small twist angles. For a tBLG at around the magic angle, there are about 12,000 atoms in a twisted moir\'e supercell. Direct simulation of a complete heterostructure requires substantial computational resources and does not yield a clear physical picture in a straightforward manner. Here we take a different approach to van der Waals heterostructure system, by a multi-scale numerical scheme as outlined in Fig. \ref{TwBLG_multiscale}. This multi-scale numerical scheme is designed to combine the accuracy of DFT calculations and the transparent physical picture of efficient low-energy $\vec{k} \cdot \vec{p}$ continuum theories, connected by a simplified tight-binding Hamiltonian for a moir\'e supercell. 
Our studies begin with the mechanical and electronic modeling for the much smaller local systems sampled from the heterostructure supercell. The full physical picture of the entire heterostructure is obtained by ``stitching'' together this local information. 

The backbone of this approach, the microscopic atomic coupling and energetics, 
are provided by accurate DFT calculations
through both the generalized stacking fault energy (GSFE), which describes 
the mechanical energy of the bilayer stacking~\cite{Carr_GSFE,GSFE_1,GSFE_2}, and Wannier orbital coupling for the electronic properties~\cite{Bilayer_TBH,TMDC_strainTBH}. 
With the GSFE, one can determine the relaxed geometry of a tBLG within continuum elasticity theory~\cite{Carr_GSFE}. The {\it ab-initio} tight-binding Hamiltonian can be constructed from the extracted Wannier couplings and applied to the relaxed geometry~\cite{Bilayer_TBH}. 
These {\it ab-initio} tight-binding Hamiltonians can be further simplified by projections onto the low-energy degrees of freedom to derive the effective $\vec{k} \cdot \vec{p}$ Hamiltonians, which are efficient ways of describing the electronic structure. Further simplifications of the $\vec{k} \cdot \vec{p}$ Hamiltonians can be carried out by projecting the lowest $n$-bands near the charge neutrality~\cite{Ashvin_10band,Carr_10band,Carr_58band}. Our method allows us to extract the relevant parameters based on well-defined assumptions and enables study of the twist angle dependence. In the following sections, we give more details for each step in this multi-scale numerical approach.  

\subsection{Mechanical Properties for Atomic Relaxation}

We start by setting the conventions for a twisted bilayer graphene crystal, and describing the atomic relaxation effects. Two sheets of monolayer graphene (denoted as L1 and L2) are stacked together, with the L2 (L1) layer rotated counterclockwise (clockwise) by $\theta/2$. This small mismatch of the crystal orientations from two sheets due to this twist angle induces a long-wavelength interference pattern in the local atomic registry, alternating between locally AA, AB and BA stacking orders. We define the origin at the center of a locally AA stacking spot, which coincides with a six-fold rotational symmetry axis at the center of a hexagon in the honeycomb structure. To simplify our discussion, here we focus on the commensurate moir\'e supercell, which is spanned by the supercell primitive vectors $\vec{R}_i (i=1, 2)$ and the associated supercell Brillouin zone with reciprocal lattice vectors $\vec{G}_i (i=1,2)$~\cite{graphene_sc_dft1}. From the perspective of each individual layer, the relative twist of L1 and L2 also rotates the reciprocal lattice vectors from each layer. The differences of the reciprocal lattice vectors from each layer give rise to the supercell reciprocal lattice vectors ($\vec{G}_i$).

Relaxed domains of locally AB/BA stacking are energetically favored in the twisted bilayer graphene crystal~\cite{Carr_GSFE,Zhang2018}. As a result of minimizing the additional energy due to the twist, the AB/BA spots are enlarged and become 
uniform stacking regions; these are separated by domain boundaries. 
Regions of AA stacking spots are reduced in size 
with the domain boundaries between neighboring AB/BA regions connecting 
the neighboring AA stacking spots. As the twist angle $\theta$ decreases, 
the area of AB stacking regions increases, but the domain wall width remains unchanged. 
This domain structure is stabilized beyond a critical twist angle~\cite{Koshino_relax,Carr_GSFE,Zhang2018}. Another feature is the puckered out-of-plane crystal relaxations~\cite{graphene_sc_dft1}. The vertical layer separation is shortest  for the AB/BA stacking regions and largest 
for the AA regions. 

To obtain the mechanical relaxation pattern of a twisted bilayer graphene, we adopted a continuum model in combination with the GSFE~\cite{Carr_GSFE,GSFE_1,GSFE_2}. 
To describe the structural deformation, at an unrelaxed position $\bm{r}$ in a supercell, 
we define the corresponding in-plane component of the displacement vector 
$\vec{U}^{(i)} (\vec{r})$ $(i = 1, 2)$ and the out-of-plane component 
$\vec{h}^{(i)} (\vec{r}) = h^{(i)} (\vec{r}) \hat{z}$ after relaxation with layer index $i$. 
The undeformed position $\vec{r}$ before relaxation is then mapped 
as $\vec{r} \rightarrow \vec{r}+\vec{U}^{(i)}(\vec{r})+h^{(i)}(\vec{r}) \hat{z}$. 
The total mechanical energy of the twisted system has two components, 
the intra- and inter-layer components. 
The intra-layer strain energy of a single layer sheet is described by a linear
 isotropic continuum approximation~\cite{Koshino_relax}:
\begin{eqnarray}
E^\mathrm{intra} (\vec{U}^{(i)}) 
= \sum_{i} \int \frac{G}{2} (\partial_x U^{(i)}_x + \partial_y U^{(i)}_y)^2  
\nonumber \\
+ \frac{K}{2} 
\left \{ (\partial_x U^{(i)}_x - \partial_y U^{(i)}_y)^2 + (\partial _x U^{(i)}_y + \partial_y U^{(i)}_x)^2
\right \} 
d^2 \vec{r},
\end{eqnarray}
where $G$ and $K$ are shear and bulk modulus of a monolayer graphene, which we take to be $G = 9.0 \, \mathrm{eV/}$\AA$^2$, $K = 13.2 \, \mathrm{eV/}$\AA$^2$. These values are obtained with DFT by isotropically straining and compressing the monolayer and performing a quadratic fitting of the ground-state energy as a function of the applied strain or shear. 


The inter-layer energy is described by the GSFE~\cite{Carr_GSFE,GSFE_1,GSFE_2}, 
denoted as $V_\mathrm{GSFE}$,
which has been employed to explain relaxation in van der Waals heterostructures~\cite{Carr_GSFE,GSFE_2}, and 
depends only on the relative stacking between two adjacent layers. 
We obtain the $V_\mathrm{GSFE} $ by applying a $9 \times 9$ grid sampling of rigid shifts to L1 in the unit cell with respect to L2 
and extract the relaxed ground state energy at each shift. 
The optimal inter-layer separation $h(\vec{r})=|h^{(1)}(\vec{r})-h^{(2)}(\vec{r})|$ is also extracted for each shifted configuration. The GSFE at any position $\vec{r}$ 
can then be expressed as a Fourier sum: 
\begin{equation} 
V^\mathrm{GSFE} (\vec{r}) = \sum_{\vec{p}_i}  V_{\vec{p}_i} \mathrm{e}^{i \vec{p}_i \cdot \vec{r}},
\end{equation}
where $\vec{p}_i$ is defined as 
$\vec{p}_i = n_1 \vec{G}_1 + n_2 \vec{G}_2$ for $n_1, n_2$ integers, and $V_{\vec{p}_i}$ is the corresponding Fourier coefficient found by fitting the ground state energy at each shift. In terms of the $V^\mathrm{GSFE}$, the inter-layer energy $E_\mathrm{inter}$ can be then written as follows for a relaxed twisted bilayer: 
\begin{equation}
E^\mathrm{inter} (\vec{U}^{(i)}) = \int V^\mathrm{GSFE} (\bm b(\vec{r}) 
+ \vec{U}^{(1)} (\vec{ r}) -  \vec{U}^{(2)} (\vec{r})) \, d^2 \vec{r},
\end{equation} 
where $\vec{b}(\vec{r})$ is the local stacking order at an atomic position $\vec{r}$, which we can take to be the distance from the given atomic position $\vec{r}$ to the position of the nearest neighbor of the same sublattice type~\cite{math_relaxation}.
Note that the $V^\mathrm{GSFE}$ is a function of the sum of the local stacking order and the displacement vectors to describe the inter-layer stacking energy after relaxation.

The total energy is the sum of the inter-layer and the intra-layer energies: 
\begin{equation}
E^\mathrm{tot} (\vec{U}^{(i)}) = E^\mathrm{intra} (\vec{U}^{(i)}) + E^\mathrm{inter} (\vec{U}^{(i)}), 
\end{equation} 
where $\vec{U}^{(1)} (\vec{r}) = - \vec{U}^{(2)} (\vec{r})$ due to the mirror symmetry between L1 and L2~\cite{Koshino_relax} (i.e. a $C_2$ rotation in three-dimensional space with an in-plane rotation axis).
We then minimize the total energy as a function of the in-plane displacement field 
$\vec{U}^{(i)}(\vec{r})$ to obtain the optimal relaxation pattern. The relaxation pattern respects the three-fold rotation symmetry and mirror symmetry of the twisted bilayer, and the functional form can be Fourier expanded: 
\begin{equation}
\vec{U}^{(i)}(\vec{r})=-i\sum_{\vec{p}_i} U_{\vec{p}_i}^{(i)} e^{i\vec{p}_i \cdot \vec{r}}, h^{(i)}(\vec{r})=h^{(i)}_0+\sum_{\vec{p}_i} h^{(i)}_{\vec{p}_i} e^{i\vec{p}_i \cdot \vec{r}},
\label{eq:mechanical_relaxation}
\end{equation} 
where symmetry requires $U^{(i)}_{\mathcal{R}_{6} \vec{p}} = \mathcal{R}_{6} U^{(i)}_{\vec{p}}$ 
and $h^{(i)}_{\mathcal{R}_{6} \vec{p}}=h^{(i)}_{\vec{p}}$ 
with $\mathcal{R}_{6}$ the $2\pi/6$ rotation matrix.

\subsection{Ab-initio Tight-Binding Hamiltonian}

Given a relaxed tBLG crystal with a commensurate supercell structure, 
the conventional Bloch theorem applies due to the presence of supercell translation symmetries. Modeling such a crystal with full DFT simulations at small twist angles is 
computationally demanding so we instead employ the  {\it ab initio} tight-binding Hamiltonian method~\cite{Bilayer_TBH,TMDC_strainTBH}. 
In this model, atomic orbital couplings are short-ranged and determined by local geometry such as the atomic registry~\cite{Bilayer_TBH}, layer separation~\cite{TWBLG_pressure}, strain~\cite{TMDC_strainTBH} and orientation in the heterostructure. 
Calculations of the much smaller aligned bilayer structure in DFT allow us to extract the relevant tight-binding parameters and their dependence on the local geometry~\cite{Bilayer_TBH,TWBLG_pressure}. 
This is based on the Wannier transformation of DFT calculations. 
The large twisted supercell structure can then be parametrized by these {\it ab initio} tight-binding Hamiltonians. We have validated the method with the full DFT simulation of a tBLG at larger twist angles~\cite{Bilayer_TBH}. 
This approach has been applied to the rigid twisted bilayers and the tBLG under external pressure~\cite{TWBLG_pressure}. The scaling with pressure is consistent with the recent experiment of the tBLG in a pressure cell~\cite{TWBLG_pressure_EXP}. 
The resulting accurate {\it ab initio} tight-binding Hamiltonians are still 
rather complicated but can be expanded to derive the simpler 
low-energy effective theories in the next section.

\subsection{Low-Energy Effective k.p Hamiltonians}

To obtain insights on the electronic structure and simpler efficient computational models,
we derive the continuum low-energy effective model based on the expansion of the  
{\it ab-initio} tight-binding Hamiltonians of a relaxed tBLG. 
We begin with a brief review of the symmetries of the crystal, the low-energy effective Hamiltonian for 
the unrelaxed twisted bilayer graphene~\cite{MacDonald_KP}, and then the generalization to a relaxed crystal.
This is followed by a numerical projection method to derive the effective models from {\it ab initio} tight-binding Hamiltonians, which is described in the following sections.

\begin{figure}[h]
  \centering
  \includegraphics[width=.5\textwidth]{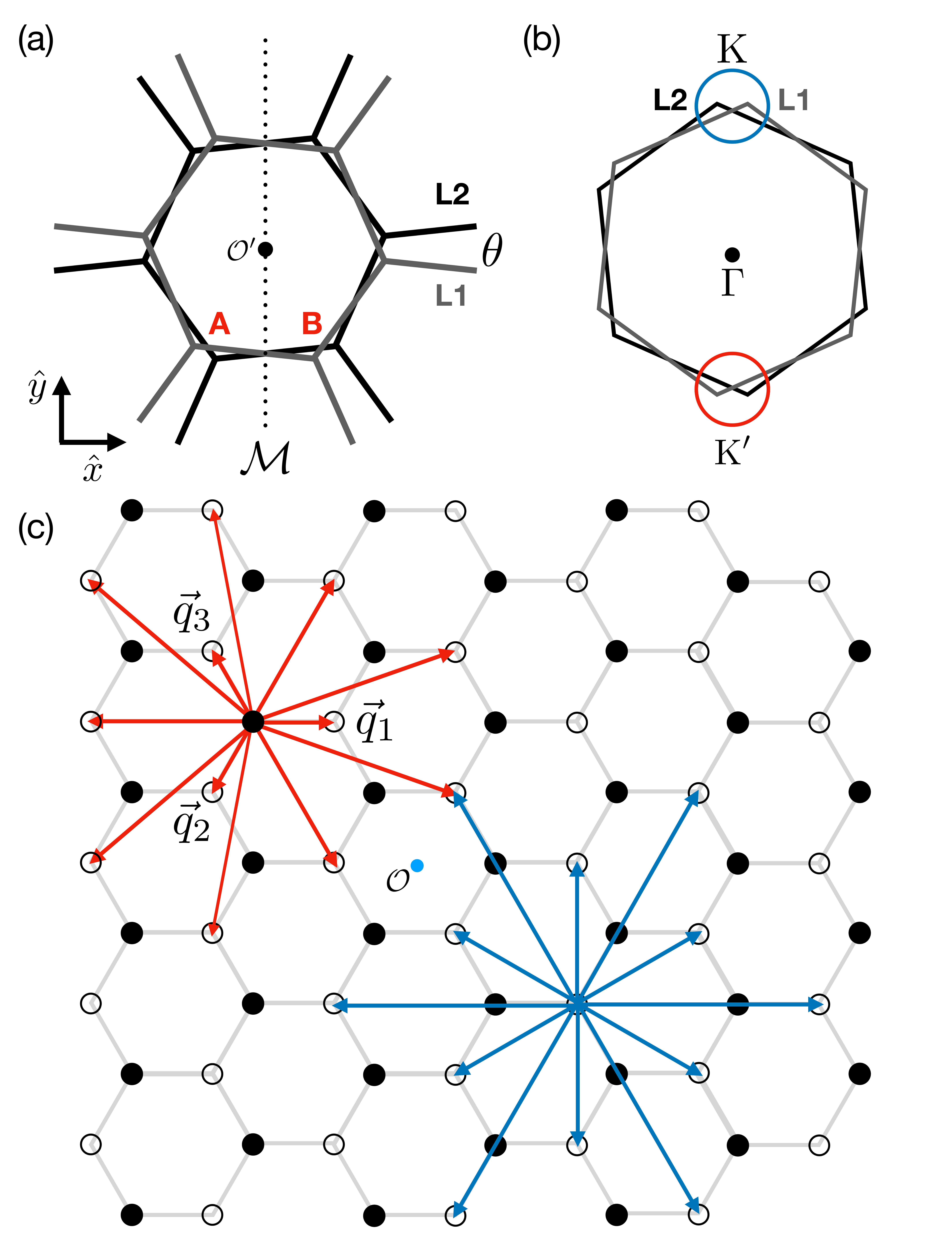}
  \caption[Conventions for a tBLG crystal and the symmetries]{Conventions for a tBLG crystal and the symmetries. (a) In the real space, the twisted bilayer is generated by the rotation around the common hexagon center from an AA-stacking bilayer unit. The symmetries are defined with respect to this hexagon centers. $\mathcal{M}$ 
is a valley-preserving mirror symmetry which is a $180^{\circ}$ rotation with respect to an 
in-plane axis shown as dotted line. 
(b) The monolayer Brillouin zones from both layers, which are rotated from each other. The effective long-wavelength theories are expanded at either the K or K' valleys. 
(c) Illustration of the low-energy $\vec{k} \cdot \vec{p}$ effective model in momentum space: 
the filled (empty) circles are momentum states from L1 (L2). 
The inter-layer (intra-layer) couplings terms with momentum $\vec{q}_i$ ($\vec{p}_i$) are represented by the red (blue) lines which connect the coupled momentum states.}
\label{TwBLG_conv}
\end{figure}

Monolayer graphene features relativistic Dirac electrons at K, K' valleys in the Brillouin zone corners. The twist angle introduces a relative displacement of Dirac cones, which originates from the same valley of each individual layer. These two copies of the Dirac electrons are then coupled through the inter-layer interaction, which is spatially varying with the moir\'e pattern. The scattering between Dirac electrons from the opposite valleys (K and K') is negligible due to the much larger scattering momentum required compared to the characteristic momentum scale of the moir\'e pattern at small twist angle. Therefore, Dirac electrons from K and K' valleys are essentially decoupled in the single particle description. Such valley protection and degeneracy, or the emergent valley $U_v(1)$ symmetry~\cite{Zou_TWBLG}, allows for the construction of effective models involving only one single K valley type with the opposite valley related by time reversal. Each copy of the effective model involves low-energy states enclosed in one circle in Fig. \ref{TwBLG_conv} (b) with the rotated monolayer Brillouin zones.

The above view of the effective $\vec{k} \cdot \vec{p}$ low-energy theory stresses its role as the expansion of the full model. On the other hand, 
symmetry considerations have been used as the guiding principle in constructing 
these low-energy models. 
The relevant symmetry operations for a tBLG effective $\vec{k} \cdot \vec{p}$ model 
include the lattice translations, $C_3$ rotations, 
valley-preserving $\mathcal{M}$ mirror symmetry and anti-unitary $C_2\mathcal{T}$ symmetry~\cite{Zou_TWBLG}. 
In our convention, the origin (rotation center) is chosen to be the hexagon center of a locally AA-stacking spot with the symmetries shown in Fig. \ref{TwBLG_conv} (a). 
Under these symmetries, the wavefunction $\Psi(\vec{r})=[\Psi_{\rm 1A}(\vec{r}), \Psi_{\rm 1B}(\vec{r}), \Psi_{\rm 2A}(\vec{r}), \Psi_{\rm 2B}(\vec{r})]$ transforms as
\begin{align}
& C_3: &  & \mathcal{M}:  &  & C_2 \mathcal{T}: & \nonumber \\
& \begin{bmatrix}
\bar{\phi} \Psi_{\rm 1A}(\mathcal{R}_3 \vec{r}) \\
\phi \Psi_{\rm 1B}(\mathcal{R}_3 \vec{r}) \\
\bar{\phi} \Psi_{\rm 2A}(\mathcal{R}_3 \vec{r}) \\
\phi \Psi_{\rm 2B}(\mathcal{R}_3 \vec{r}) 
\end{bmatrix}, &
& \begin{bmatrix}
\Psi_{\rm 2B}(\mathcal{M} \vec{r}) \\
\Psi_{\rm 2A}(\mathcal{M} \vec{r}) \\
\Psi_{\rm 1B}(\mathcal{M} \vec{r}) \\
\Psi_{\rm 1A}(\mathcal{M} \vec{r}) 
\end{bmatrix}, &
& \begin{bmatrix}
\Psi^*_{\rm 1B}(-\vec{r}) \\
\Psi^*_{\rm 1A}(-\vec{r}) \\
\Psi^*_{\rm 2B}(-\vec{r}) \\
\Psi^*_{\rm 2A}(-\vec{r}),
\end{bmatrix} 
\end{align} 
where $\phi=e^{i2\pi/3}$ ($\bar{\phi}=e^{-i2\pi/3}$), $\mathcal{R}_3$ rotates 
the vector $\vec{r}$ clockwise by $2\pi/3$ and $\mathcal{M}$ 
flips the $x$ coordinate of the vector $\vec{r}$ 
as in Fig. \ref{TwBLG_conv} (a). The crystal relaxation retains these relevant symmetries.

Such an effective low-energy theory expanded around the K valley has already been derived
for the {\em unrelaxed} case of tBLG~\cite{MacDonald_KP,graphene_sc_kp1,graphene_sc_kp2,graphene_sc_kp3}. 
Since here we want to eventually include the electronic effects of relaxation, we augment this 
model to include additional terms that can capture these effects.
With the gauge convention chosen such that the origin coincides with a locally AA-stacking spot, the augmented low-energy Hamiltonian takes the form:
\begin{equation}
\label{eq:eff_relax_kp}
\tilde{H}_{\rm K}=\begin{bmatrix}
H^{(1)}_D(\vec{k}) +A^{(1)}(\vec{r}) & \tilde{T}^\dagger(\vec{r}) \\
\tilde{T}(\vec{r}) & H^{(2)}_D(\vec{k}) +A^{(2)}(\vec{r}),
\end{bmatrix}
\end{equation} 
where $H^{(i)}_D(\vec{k})$ is the Dirac Hamiltonian for each individual layer ($i=1,2$), and
$\tilde{T}(\vec{r})$ the inter-layer coupling matrix, which varies with the spatial moir\'e pattern. The Dirac Hamiltonian is given by:
\begin{equation}
H_D(\vec{k}) =v_F\begin{bmatrix}
0 &  -i k_+ \\
i k_-   &   0
\end{bmatrix},
\end{equation} with additional Pauli matrix rotation terms to account for the small twist angle.


In the unrelaxed case, the terms $A^{(i)}(\vec{r}), i=1,2$ are set to zero, and the inter-layer coupling $\tilde{T}(\vec{r})$ is simplified to:
\begin{equation}
T_{\alpha \beta}(\vec{r}) =\omega_0 \begin{bmatrix}
T_0(\vec{r}) & T_+(\vec{r}) \\
T_-(\vec{r}) & T_0(\vec{r})
\end{bmatrix},
\end{equation}
where $T_{\alpha \beta}(\vec{r})$ specifies the coupling between various momentum states from the sublattice $\beta$ of L1 to sublattice $\alpha$ of L2, $\alpha, \beta={\rm A}, {\rm B}$.
In the above equations, the various symbols that 
appear have the following meaning:
$k_{\pm}=k_x\pm i k_y$, $T_0(\vec{r})=e^{i \vec{q}_1 \cdot \vec{r}}+e^{i \vec{q}_2 \cdot \vec{r}}+e^{i \vec{q}_3\cdot \vec{r}}$, $T_-(\vec{r})=e^{i \vec{q}_1 \cdot \vec{r}}+\phi e^{i \vec{q}_2 \cdot \vec{r}}+ \bar{\phi} e^{i \vec{q}_3\cdot \vec{r}}$, $T_+(\vec{r})=e^{i \vec{q}_1 \cdot \vec{r}}+\bar{\phi} e^{i \vec{q}_2 \cdot \vec{r}}+ \phi e^{i \vec{q}_3\cdot \vec{r}}$. $\vec{q}_1=k_D \hat{x}$, $\vec{q}_2=k_D(-\hat{x}-\sqrt{3}\hat{y})/2$ and $\vec{q}_3=k_D(-\hat{x}+\sqrt{3}\hat{y})/2$ with $k_D=(8\pi/3a_G)\sin(\theta/2)$ 
and $a_G$ being the graphene lattice constant, as shown in Fig. \ref{TwBLG_conv} (c). 
The inter-layer coupling strength is $\omega_0 \approx $ 110 meV. The moir\'e supercell has reciprocal lattice vectors $\vec{G}_1=\vec{q}_1-\vec{q}_2$ and $\vec{G}_2=\vec{q}_3-\vec{q}_2$. The Hamiltonian can be shown to preserve the symmetries above. One interesting observation is that the coupling $T(\vec{r})$ is not periodic under moir\'e supercell translations~\cite{MacDonald_butterfly} due to the field expansion gauge choice around
the Dirac point of each individual layer. The two shifted Dirac cones are connected by $\vec{q}_1$ as in Fig. \ref{TwBLG_conv} (b) and (c). 
We will re-examine this labeling of momentum states which would facilitate the projection 
from the full tight-binding Hamiltonian.

To determine the set of coupled momentum states, we first look for the momentum states in each layer that are folded onto the same supercell momentum label. With $\vec{k}_i$ being the momentum measured from the 
${\rm K}^{(i)}$ valley, this requires ${\rm K}^{(1)} + \vec{k}_1 = {\rm K}^{(2)}  + \vec{k}_2 +\vec{G}$, where $\vec{G}$ is a reciprocal lattice vector of the moir\'e supercell spanned by $\vec{G}_i$. The set of $\vec{k}_i$ momentum states of each layer (filled and empty circles) that are folded to the supercell $\Gamma$ point are shown in Fig. \ref{TwBLG_conv} (c), which is a bipartite lattice in momentum space. 
In the absence of the deformation from lattice relaxation, the exact translation symmetry within each single layer eliminates the direct intra-layer coupling terms between momentum states. This leaves only the inter-layer coupling $T(\vec{r})$ to be determined. The inter-layer coupling $T(\vec{r})$ describes the coupling between states at momentum 
$\vec{k}_2 - \vec{k}_1 = ({\rm K}^{(1)} - {\rm K}^{(2)}) -\vec{G}$,  which does not belong to moir\'e supercell reciprocal lattice vectors. In a rigid tBLG, the dominant contributions in $T(\vec{r})$ can be derived from the orbital coupling in the microscopic tight-binding Hamiltonian~\cite{MacDonald_KP}. The low-energy Hamiltonian 
can be viewed as a momentum lattice with only nearest neighbor couplings. The $2 \times 2$ coupling matrices to the three nearest neighbors direction $\vec{q}_i$ are:
\begin{equation}
T_1= \omega_0 \begin{bmatrix}
1 & 1\\
1 & 1
\end{bmatrix}, \; \; 
 T_2= \omega_0 \begin{bmatrix}
1 &   \bar{\phi} \\
\phi  &  1 
\end{bmatrix}, \; \; 
T_3= \omega_0 \begin{bmatrix}
1 &  \phi  \\
\bar{\phi}  &  1 
\end{bmatrix}.
\end{equation} 
A momentum cutoff is also imposed for the momentum lattice within the linear Dirac cone regime.

To generalize the above effective Hamiltonian 
for a twisted bilayer crystal with atomic relaxation within the plane (in-plane strain) 
as well as height variations (out-of-plane strain), 
we start with a deformed monolayer graphene with a strain field $\vec{U}^{(i)}(\vec{r})$, which describes the displacement vector of the underlying constituent atoms, 
moving the atom to a new position $\vec{r} \rightarrow \vec{r}+\vec{U}^{(i)}(\vec{r})+h^{(i)}(\vec{r})\hat{z}$ on the $i$-th layer. 
The in-plane strain is known to appear in the low-energy theory as the pseudo-gauge field with $A^{(i)}(\vec{r})$ a $2 \times 2$ matrix which couples with the Dirac Hamiltonian $H_D^{(i)}(\vec{k})+A^{(i)}(\vec{r})$~\cite{eff_graphene_strain}. 
The isotropic part of the strain $(\partial_x U^{(i)}_{x}+\partial_y U^{(i)}_{y})$ 
or the deformation potential contributes to the diagonal part of $A^{(i)}(\vec{r})$, 
while the off-diagonal elements are given by the coupling 
to $(\partial_x U^{(i)}_{x}-\partial_y U^{(i)}_{y})$ and $(\partial_x U^{(i)}_{y}+\partial_y U^{(i)}_{x})$ strain components dictated by symmetry. 
The spatially varying $A^{(i)}(\vec{r})$ matrix introduced by the atomic deformation induces coupling between different momentum states within the same single layer. The $A^{(i)}(\vec{r})$ matrix can be decomposed into Fourier components at moir\'e supercell reciprocal lattice vectors.
Strain perturbations are also known to renormalize the local Fermi velocity of the Dirac electron and give anisotropic velocity corrections~\cite{eff_graphene_strain}. Here, we only retain the contributions in the form of the $A^{(i)}(\vec{r})$ matrix, which causes scatterings between the momentum states. For a twisted bilayer, the mirror symmetry will relate the $A^{(i)}(\vec{r})$ between the two layers.

Regarding the inter-layer coupling, the deformation field $\vec{U}^{(i)}(\vec{r})+h^{(i)}(\vec{r})\hat{z}$ from relaxation causes additional relative shifts in the local atomic registry, which modifies the $T(\vec{r})$ coupling matrix. The height variations in the moir\'e supercell weaken the inter-layer coupling in the locally AA stacking region (due to its larger vertical separation) and enhance the coupling in the enlarged  AB/BA domains. This causes the off-diagonal coupling constants in  $\tilde{T}(\vec{r})$ to be larger than the diagonal components. The dominant $T_1$ above with atomic relaxation becomes $\omega_0 \sigma_0 +\omega_1 \sigma_x$ with $\omega_0 < \omega_1$. These corrections are relevant for the single-particle gaps above and below the flat bands near the magic angle. Furthermore, the domain line formation and the finer structure from the atomic relaxation enhances the scattering with higher momentum components. The dependence of the coupling parameters on the twist angle will be discussed later.



In short summary, the intra-layer deformation of each of the two layers introduces the non-zero scattering terms, $A^{(i)}(\vec{r})$, in Eq. (\ref{eq:eff_relax_kp}). 
The Dirac electrons from the two layers are then coupled through the inter-layer $\tilde{T}(\vec{r})$ term, which is also modified by the atomic relaxation. Later, we will include another correction term to the inter-layer coupling that involves momentum dependence of the scattered momentum states.
 

The effects of various coupling terms here can also be visualized from the momentum space representation of the Hamiltonian. For simplicity, we first focus on the supercell $\Gamma$ point in momentum space. The relevant momentum states from both layers are represented by the circles in Fig. \ref{TwBLG_conv} (c). The inter-layer coupling from $\tilde{T}(\vec{r})$ links the states from both layers, while the intra-layer $A^{(i)}(\vec{r})$ connects states reside within the same layer. We can expand these coupling terms into the Fourier momentum components
\begin{equation}
A^{(i)}(\vec{r}) = \sum_{\vec{p}_i} A^{(i)}_{\vec{p}_i} e^{i \vec{p}_i \cdot \vec{r}}, \; \; 
\tilde{T}(\vec{r}) = \sum_{\vec{q}_i} \tilde{T}_{\vec{q}_i} e^{i \vec{q}_i \cdot \vec{r}},
\end{equation}
with the twelve $\vec{q}_i$ ($\vec{p}_i$) momentum vectors illustrated in Fig. \ref{TwBLG_conv} (c). The $\vec{p}_i$ vectors are the reciprocal lattice vectors of the moir\'e supercell, $\vec{p}_i=n_1 \vec{G}_1+n_2 \vec{G}_2$ and the general $\vec{q}_i$ vectors are $ ({\rm K}^{(1)} - {\rm K}^{(2)}) -\vec{G}^{'}$. These correction terms introduce further neighbor couplings in the momentum lattice to the model as in Fig. \ref{TwBLG_conv} (c).

We now discuss how the symmetries in the low-energy theory constrain the forms of the coupling terms. For the $C_3$ and $C_2 \mathcal{T}$ symmetries, which do not flip the layer indices, 
we let $F(\vec{r})$ represent either of $A^{(i)}(\vec{r})$ or $\tilde{T}(\vec{r})$ $2 \times 2$ matrices. 
These symmetries dictate
\begin{align}
 C_3 &:& e^{i(2\pi/3)\sigma_z}  F(\mathcal{R}_3^{-1} \vec{r})  e^{-i(2\pi/3)\sigma_z}  = F(\vec{r}), \\
C_2 \mathcal{T} & : & \sigma_x F^*(-\vec{r}) \sigma_x = F(\vec{r}),
\end{align} ($\mathcal{R}_3^{-1}$ rotates $\vec{r}$ counterclockwise) with the Pauli matrices $\vec{\sigma}=(\sigma_x,\sigma_y,\sigma_z)$ acting on the sublattice degree of freedom. As for the $\mathcal{M}$ mirror symmetry that flips the layer, we can derive the following relations
\begin{align}
\mathcal{M} & : & 
\sigma_x \tilde{T}(\mathcal{M}\vec{r}) \sigma_x = \tilde{T}^\dagger (\vec{r}), \; \; 
\sigma_x A^{(1)}(\mathcal{M}\vec{r}) \sigma_x =A^{(2)} (\vec{r}),
\end{align} 
and similarly for $A^{(1)} \leftrightarrow A^{(2)}$ in the last equation. When these $2 \times 2$ matrices are decomposed into Fourier momentum components, these symmetry conditions yield the following equivalent constraints on the expanded matrix components:
\begin{align}
 C_3 & : & e^{i\frac{2\pi}{3}\sigma_z}  F_{\mathcal{R}_3^{-1}\vec{g}_i} e^{-i\frac{2\pi}{3}\sigma_z} = F_{\vec{g}_i}, \\
 C_2 \mathcal{T} & : & \sigma_x F^*_{\vec{g}_i}   \sigma_x =  F_{\vec{g}_i} , \\
 \mathcal{M} & : & \sigma_x A^{(1)}_{\mathcal{M} \vec{p}_i} \sigma_x = A^{(2)}_{ \vec{p}_i}, \;
 \sigma_x  \tilde{T}_{-\mathcal{M} \vec{q}_i}  \sigma_x = \tilde{T}_{\vec{q}_i}^{\dagger}.
\end{align} 
In the original effective model~\cite{MacDonald_KP}, only the first three $\vec{q}_i$'s (connections to the nearest neighbors) from $\tilde{T}_{\vec{q}_i}$ are included. Even though these three $\vec{q}_i$'s are shown to be the dominant contribution in an unrelaxed twisted bilayer, relaxation would not only modify the matrix elements but also enhance the higher momentum components. 


\subsection{Low-Energy Expansion based on Ab-initio Tight-Binding Hamiltonian}
Here, we summarize the procedure to  extract these relevant matrix elements for both the pseudo-gauge field $A^{(i)}(\vec{r})$ and inter-layer coupling $\tilde{T}(\vec{r})$ in the effective low-energy theory, derived from an {\it ab initio} tight-binding Hamiltonian of relaxed twisted bilayer graphene.

(i) For a relaxed commensurate twisted bilayer with atomic relaxation, the supercell is spanned by the translation vectors $\vec{R}_i$. Two integers $M$ and $N$ are used to specify the twist angle~\cite{graphene_sc_dft1}. The origin is chosen to be at the center of the hexagon as in Fig.~\ref{TwBLG_conv} (a). We adopt the series $M=N+1$ commensurate supercells which have vanishing twist angles with increasing $M$ and exactly one AA-stacking region in a supercell unit. For an unrelaxed twisted bilayer, these specify all the atomic positions $\vec{r}_l$ in a supercell. The relaxation moves an atom to a new position by $\vec{r}_l \rightarrow \vec{r}_l  + \vec{U}^{(i)}(\vec{r}_l) + h^{(i)}(\vec{r}_l) \hat{z}$ and modifies the coupling strength between atomic pairs. Given a supercell momentum $\vec{k}_{\rm SC}$, we define the Bloch wave as
\begin{equation}
| \Psi_l (\vec{k}_{\rm SC})  \rangle=\frac{1}{\sqrt{N_{\rm SC}}} \sum_{\vec{R}} e^{i\vec{k}_{\rm SC} \cdot (\vec{r}_l+\vec{R})} | \vec{r}_l+\vec{R} \rangle,
\end{equation} 
where $N_{\rm SC}$ is the total number of supercells. Note that the original unrelaxed position $\vec{r}_l+\vec{R}$ is used in the definition even though the actual relaxed position is $\vec{r}_l+\vec{U}^{(i)}(\vec{r}_l) + h^{(i)}(\vec{r}_l) \hat{z} +\vec{R}$. The Hamiltonian $H^{\rm TBH}(\vec{k}_{\rm SC})$ in the supercell reciprocal space can be derived as: 
\begin{eqnarray}
H^{\rm TBH}_{lm}(\vec{k}_{\rm SC})=\sum_{\vec{R},\vec{R}'} t_{\vec{r}_l+\vec{U}^{(i)}(\vec{r}_l)+\vec{R},\vec{r}_m+\vec{U}^{(j)}(\vec{r}_m)+\vec{R}'} \\ 
\times e^{-i\vec{k}_{\rm SC} \cdot (\vec{r}_l+\vec{R}-\vec{r}_m-\vec{R}')}.\nonumber
\end{eqnarray}

(ii) The relevant states at low-energy are the Bloch waves near the K valley of both layers.
\begin{align}
&|\Phi_{\alpha} ({\rm K}^{(i)} + \vec{k}^{(i)})  \rangle  \nonumber \\
&= \frac{1}{\sqrt{N_{\rm SC} N_{\alpha}}} \sum_{\vec{R}} \sum_{\vec{r}_l \in \alpha} e^{i ({\rm K}^{(i)} + \vec{k}^{(i)}) \cdot (\vec{r}_l+\vec{R} )} | \vec{r}_l+\vec{R} \rangle \nonumber \\
& = \frac{1}{\sqrt{N_{\alpha}}}  \sum_{\vec{r}_l \in \alpha} e^{i \vec{G}' \cdot \vec{r}_l } | \Psi_l (\vec{k}_{\rm SC})  \rangle,
\end{align} 
with $i=1,2$ being the layer index, $\alpha=A, B$ the sublattice index, 
and $\vec{k}^{(i)}$ is the momentum measured relative to ${\rm K}^{(i)}$. 
These states are defined with the unrelaxed original positions and momentum labeling from graphene primitive unit cell translations. 
To be compatible with the translational symmetry of the above Hamiltonian $H^{\rm TBH}(\vec{k}_{\rm SC})$, 
the momentum states must be such that they can be 
folded to the supercell $\vec{k}_{\rm SC}$ point in momentum space. 
This yields the condition ${\rm K}^{(i)}+\vec{k}^{(i)}=\vec{G}'+\vec{k}_{\rm SC}$ with $\vec{G}'$ being a supercell reciprocal lattice vector. 
Given a supercell momentum $\vec{k}_{\rm SC}$, we can then construct a set of Bloch plane wave states for both layers to sample the low-energy sector of the Hamiltonian. The unrelaxed $\vec{r}_l+\vec{R}$ positions of atoms used in the Bloch phase factors ensure the orthogonality of the projection basis states, for a crystal with or without atomic relaxation. The state 
$|\Phi_{\alpha} ({\rm K}^{(i)} + \vec{k}^{(i)})  \rangle$ in the tight-binding orbital picture is mapped to the plane wave state on $i$-th layer with momentum $\vec{k}^{(i)}$ in the low-energy $\vec{k} \cdot \vec{p}$ expansion. We will use this picture to bridge the tight-binding and low-energy $\vec{k} \cdot \vec{p}$ pictures.

\begin{widetext}

\begin{figure}[h]
  \centering
  \includegraphics[width=1.0\textwidth]{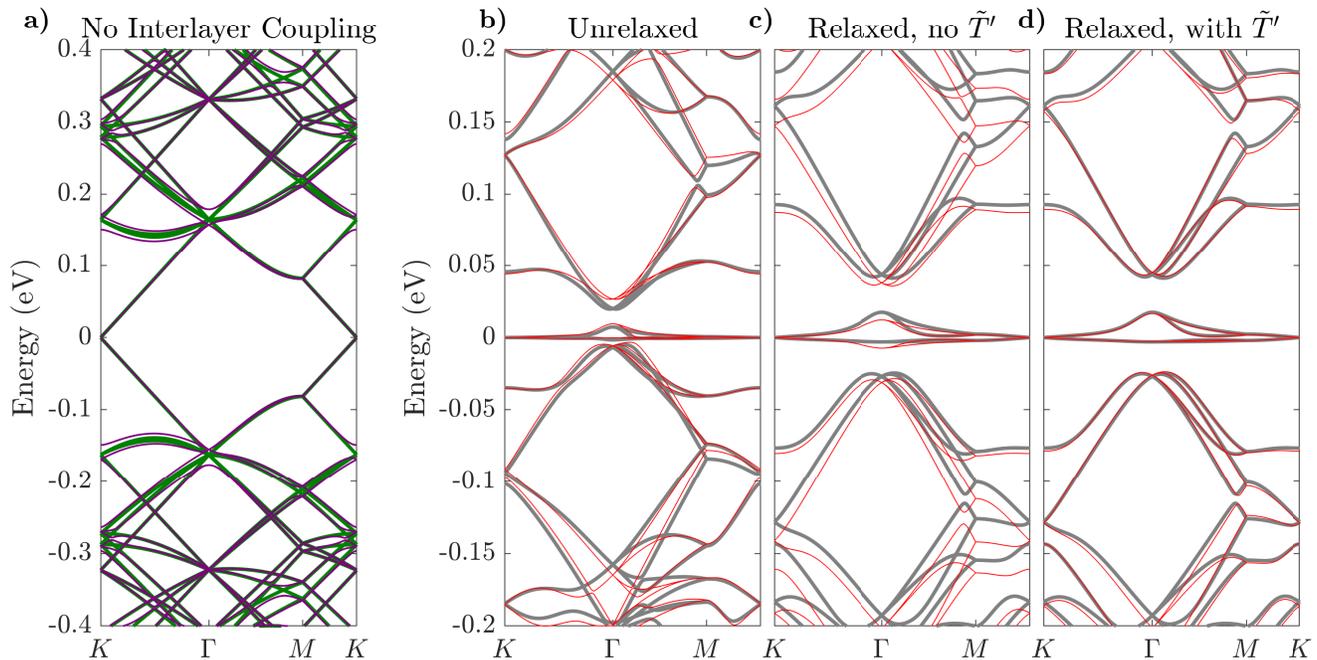}
  \caption{
  \textbf{(a)} A comparison of the band structures from decoupled bilayers, one without relaxation (thick green lines) and one with relaxation (purple lines).
  The in-plane and out-of-plane relaxation can open 
band gaps at high-symmetry points.
  \textbf{(b-d)} Band structures benchmarked for effective $\vec{k} \cdot \vec{p}$ bilayer Hamiltonians at $\theta = 1.05^\circ$ with a momentum lattice cutoff radius of $6 |\vec{G}|$.
  The full tight-binding reference bands 
are in thick grey lines and the $\vec{k} \cdot \vec{p}$ bands are in red.
  We show examples for both the unrelaxed bandstructure \textbf{(b)} and the relaxed bandstructure, \textbf{(c)} without momentum-dependent and \textbf{(d)} with 
momentum-dependent $\tilde{T}'(\vec{r},\hat{k}_{\pm})$ terms in Eq. (\ref{eq:kdep_T}).
}
\label{fig:kp_benchmark}
\end{figure}

\end{widetext}

(iii) Having established
 the tight-binding Hamiltonian and the relevant low-energy states $|\Phi_{\alpha} ({\rm K}^{(i)} + \vec{k}^{(i)})  \rangle$, 
we can project out the coupling matrix elements in $A_{\vec{p}_j}^{(i)}$ and $\tilde{T}_{\vec{q}_j}$ for the low-energy Hamiltonian. For example, the intra-layer scattering terms from $A^{(i)}_{\alpha \beta}$ with momentum $\vec{p}_j$ can be inferred from $ \langle \Phi_{\alpha} ({\rm K}^{(i)} + \vec{k}^{(i)}+\vec{p}_j) | H^{\rm TBH}(\vec{k}_{\rm SC})  | \Phi_{\beta} ({\rm K}^{(i)} + \vec{k}^{(i)})\rangle $ with ${\rm K}^{(i)} + \vec{k}$ and ${\rm K}^{(i)} + \vec{k}+\vec{p}_j$ both folded to $\vec{k}_{\rm SC}$. 
The inter-layer coupling $\tilde{T}_{\alpha \beta}$ at momentum $\vec{q}_j$ can be obtained from $ \langle \Phi_{\alpha} ({\rm K}^{(2)} + \vec{k}+\vec{q}_j) | H^{\rm TBH}({\vec{k}_{\rm SC})}  | \Phi_{\beta} ({\rm K}^{(1)} + \vec{k})\rangle$ with the momentum transfer $\vec{q}_j$ and compatible momentum states. 
Numerically, the coupling matrix is obtained from the average of sampling $\vec{k}_{\rm SC}$ and $\vec{k}^{(i)}$ momenta near the K valley that are scattered into new states. 
The reconstructed low-energy $\vec{k} \cdot \vec{p}$ Hamiltonian captures most essential features of the tight-binding electronic band structure as shown in Fig. \ref{fig:kp_benchmark} (c). To further improve the particle-hole asymmetry of the bands, 
we identify the relevant terms to be added to the low-energy theory in the next section.

\subsection{Momentum-Dependent Inter-layer Coupling}

Experimentally, it is interesting to investigate whether the tBLG is particle-hole symmetric or not under electrical gating and the implications for the correlated insulating and superconducting phases. For the electronic structures, higher order terms in the Dirac Hamiltonian, Pauli matrix rotations, due to the twist angle~\cite{Bernevig_TWBLG} and electric potential from doping~\cite{Guinea_TWBLG} are known to give rise to particle-hole asymmetric bands. In our model, we have included the Pauli matrix rotations but they are not adequate
to fully capture the asymmetric bands in the tight-binding calculations as shows in Fig. \ref{fig:kp_benchmark} (c). The projection method introduced above from the full tight-binding Hamiltonian enables us to systematically extract and identify various terms and approximations. In the above $\vec{k} \cdot \vec{p}$ low-energy Hamiltonian, the inter-layer couplings between $|\Phi_{\alpha}^{(2)} ({\rm K}^{(2)} + \vec{k}+\vec{q}_j) \rangle$ and $ |\Phi_{\beta}^{(1)} ({\rm K}^{(1)} + \vec{k} ) \rangle$ states are assumed to only depend on the momentum transfer $\vec{q}_j$. However, there is no protecting symmetry for this property and the coupling constants could generally depend on the momentum $\vec{k}$ as well. This dependence is explicitly seen in our numerical projection method. To elucidate these momentum-dependent correction terms, we focus here on the first three dominant $\vec{q}_j$ contributions,
which gives the following form
\begin{equation}
\label{eq:kdep_T}
\begin{split}
\tilde{T}'(\vec{r},\hat{k}_{\pm})=&\frac{i}{2}\begin{bmatrix}
\lambda_1 \big\{ T_-(\vec{r}), \hat{k}_+\big\} & \lambda_3 \big\{ T_0(\vec{r}), \hat{k}_+ \big\}  \\
\lambda_2 \big\{ T_+(\vec{r}), \hat{k}_+ \big\}  &  \lambda_1 \big\{ T_-(\vec{r}), \hat{k}_+ \big\} \end{bmatrix} \\
&-\frac{i}{2}\begin{bmatrix}
\lambda_1 \big\{ T_+(\vec{r}), \hat{k}_- \big\} & \lambda_2 \big\{ T_-(\vec{r}), \hat{k}_- \big\}  \\
\lambda_3 \big\{ T_0(\vec{r}), \hat{k}_- \big\}  &  \lambda_1 \big\{ T_+(\vec{r}), \hat{k}_- \big\} 
\end{bmatrix},
\end{split}
\end{equation} 
where 
$\lambda_2 \approx 2\lambda_1 \approx 0.18$ eV$\cdot$\AA, $\lambda_3 \approx 0$ around $\theta=1^\circ$ and the
anti-commutator form for non-commuting operators. We have added this leading momentum-dependent scattering terms in the inter-layer coupling, $\tilde{T}(\vec{r})+\tilde{T}'(\vec{r},\hat{k}_{\pm})$, and they indeed 
can capture the particle-hole asymmetry as shown in Fig. \ref{fig:kp_benchmark} (d). 
The projected low-energy bands are in good agreement with the full {\it ab initio} tight-binding electronic structure.



\section{Numerical Results For Angular Dependence}

The effective models presented above 
are derived for twist angles in the range of  $0.7^\circ$ to $4^\circ$. A series of commensurate $(M,N)$ supercells with $M=N+1$ are used to extract the model parameters in this range. 
For the general twist angles, the parameters can be obtained by interpolating the results of the commensurate cases. 
The numerical results are presented in their Fourier component version. 
There are two sets of crystal momenta involved in the expansions. 
The first is denoted as $\vec{p}_i=n_1 \vec{G}_1+n_2 \vec{G}_2$ which can be expressed 
by the moir\'e supercell reciprocal lattice vectors $\vec{G}_i$. 
The other second can be written as $\vec{q}_i$ as in Fig. 2. 
The differences between $\vec{q}_i$ vectors are reciprocal lattice vectors 
of the moir\'e supercell. 
Hence, any $\vec{q}_i$ can be written as $\vec{q}_i=\vec{q}_1+m_1^i \vec{G}_1+m_2^i \vec{G}_2$.
The $\vec{q}_i$ are the basis for the inter-layer coupling expansion, 
while the $\vec{p}_i$ are for the intra-layer expansion (strain).
Terms evaluated from momenta of identical magnitude tend to only differ by a complex phase dictated by symmetries.
For this reason, we organize the $\vec{p}$ and $\vec{q}$ 
dependent terms by ``shells'' of equidistant momenta.
For example, in Fig. \ref{TwBLG_conv}, there are three shells of $\vec{q}$ (in red) 
and two shells of $\vec{p}$ (in blue).
In summarizing the strength of terms in the expansion, 
it is more convenient to compare the average magnitudes of different shells instead 
of the complex values of individual momenta, 
so we label these terms as $T_{s_j}$ and $A_{s_j}$ instead of $T_{\vec{q}_i}$ and $A^{(i)}_{\vec{p}_i}$

\begin{figure}[h]
\includegraphics[width=1.0\linewidth]{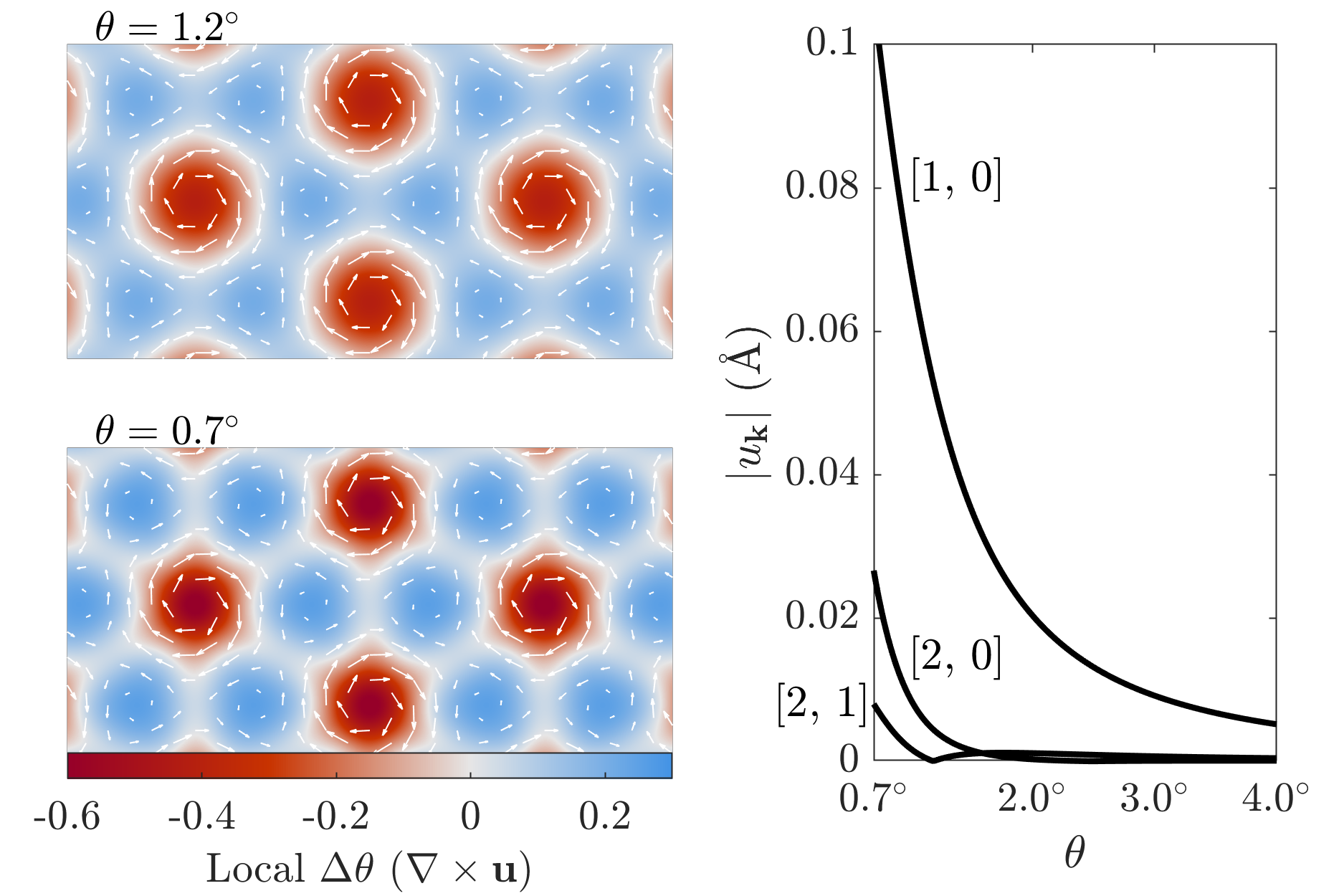}
\caption{
  (Left) Relaxation patterns of L2 for bilayers with twist angles $\theta = 1.2^\circ$ and $0.7^\circ$ from the continuum model.
  The local relaxation is given by the white arrows, and the curl of the relaxation (local change in twisting angle) is given by color.
  (Right) Twist-angle dependence of the relaxation coefficients $u_{\vec{p}_i}$, 
where the label $[n_1, n_2]$ corresponds to $\vec{p}_i = n_1 \vec{G}_1 + n_2 \vec{G}_2$.
}
\label{fig:relaxations}
\end{figure}

We begin with the mechanical properties of the relaxed twisted bilayer graphene. 
The competition between the stacking energy and the strain energy will depend on the twist angle which determines the moir\'e length scale.
The atomic structure will tend to reduce to the area of the unfavorable $AA$ stacking, while increasing the area of the favorable Bernal stacking ($AB$ and $BA$).
The displacement field in Eq.~(\ref{eq:mechanical_relaxation}) can be expanded into the Fourier components at momenta $\vec{p}_i$. 
A real-space image of the atomic relaxation is provided in Fig. \ref{fig:relaxations}, 
along with the angle-dependent magnitudes of the representative $\vec{p}_i$ from the first three shells.

\begin{figure}[h]
  \centering
  \includegraphics[width=1.0\linewidth]{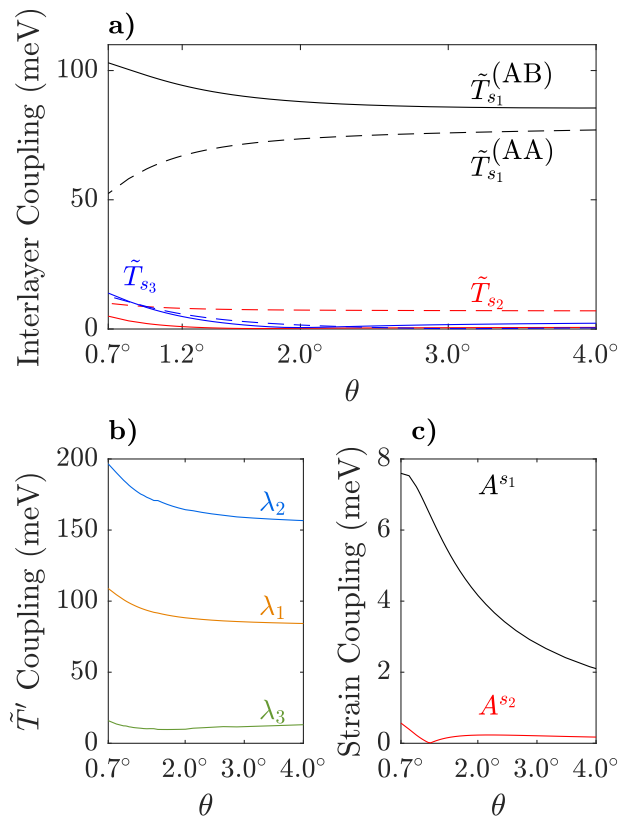}
  \caption{Twist-angle dependence of various parameters in the $\vec{k} \cdot \vec{p}$ model.
  \textbf{(a)} Magnitudes of inter-layer coupling terms $\tilde{T}_i$ for the three nearest shells ($s_j$) of couplings (red lines in Fig. \ref{TwBLG_conv} (c).). 
  The off-diagonal elements of the $\tilde{T}$ matrices (AB) are 
shown by solid lines, while the diagonal elements (AA) by dashed lines.
  \textbf{(b)} The three $\lambda_i$ momentum-dependent parameters.
  \textbf{(c)} Magnitudes of the intra-layer coupling terms, ${A}^{(i)}$ for the two nearest shells ($s_j$) of couplings (blue lines in Fig. \ref{TwBLG_conv}(c).).
  }
\label{fig:kp_terms}
\end{figure}

The atomic deformation induces a pseudo-gauge field coupled to the Dirac electrons in each layer.
The pseudo-gauge can also be Fourier expanded at $\vec{p}_i$ momenta, and then easily included in the $\vec{k} \cdot \vec{p}$ Hamiltonian.
The crystal relaxation also modifies the inter-layer coupling.
Both the $\tilde{T}(\vec{r})$ and the momentum-dependent corrections $\tilde{T}'(\vec{r})$ are affected.
The diagonal inter-layer coupling term (labeled $\tilde{T}^{(AA)}$) is generally smaller than the off-diagonal coupling (labeled $\tilde{T}^{(AB)}$), as the $AA$ stacking has a larger inter-layer distance (and thus smaller electronic coupling) than the $AB$ and $BA$ regions.
The $AA$ coupling is further reduced at small angles by the reduction of the overall $AA$ stacking area, while the $AB$ coupling is increased.
The dependence of all electronic terms on twist angle 
for the relaxed tBLG system is given in Fig. \ref{fig:kp_terms}.

\begin{figure}[h]
  \centering
  \includegraphics[width=1.0\linewidth]{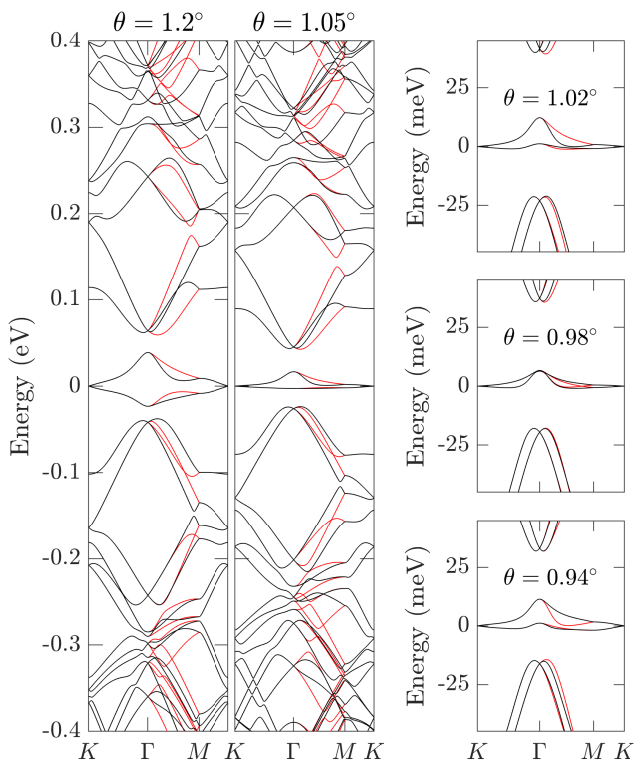}
  \caption{Twist-angle dependence of the low energy band structure of relaxed tBLG, calculated using the relaxed $k \cdot p$ model. The black lines are for an expansion around the monolayer K valley, while the red lines are for the monolayer K' valley. 
At $\theta = 0.98^\circ$, the upper and lower flat bands at the $\Gamma$ point pass smoothly through one another due to opposite mirror symmetry eigenvalues~\cite{Carr_10band}.}
\label{fig:example_bands}
\end{figure}

Using the fitted $\vec{k} \cdot \vec{p}$ parameters, band structures can be calculated for tBLG systems at an arbitrary twist angle (Fig.~\ref{fig:example_bands}).
These band structures are essentially identical to those of the full tight-binding model (see Fig.~\ref{fig:kp_benchmark}d).
Our model predicts an inversion of the two low-energy bands at $0.98^\circ$, but has nearly flat bands in a range of angles from $0.95^\circ$ to $1.05^\circ$.
For further analysis and discussion of the low-energy electronic structure and its angle-dependence we defer to a companion work, Ref. \onlinecite{Carr_10band}.

\section{CONCLUSION}
In this work, we adopted a systematic multi-scale approach to the modeling of tBLG with an angle dependence and relaxation effects included. 
Starting with DFT calculations, the mechanical and electronic properties are extracted to model the relaxed geometry 
through {\it ab-initio} tight-binding Hamiltonians. 
We then simplify the models by low-energy $\vec{k} \cdot \vec{p}$ expansions to capture the essential electronic features of the derived {\it ab initio} tight-binding Hamiltonians~\cite{Carr_10band}. 
Our approach provides an unified and coherent method to derive and simplify the modeling process while preserving the important physics, 
which paves the way for formulating interacting-electron theories (The model scripts are available at \url{https://github.com/stcarr/kp_tblg}).

There are also other factors that may affect the electronic structure of tBLG
that are not included in our current modeling.
Such factors are the hBN substrate effects on tBLG crystal relaxation and electronic couplings, screening effects from the substrate used~\cite{Louie_2d_screening1}, electrical gating, 
and
self-consistent electric potential from the electron doping~\cite{Guinea_TWBLG}. 
Disorder and additional strain variations~\cite{Zhen_strain} across the samples
introduced by the fabrication process might also complicate the interpretation of 
the observed behavior~\cite{TWBLG_STM1}. 
Our multi-scale method provides a framework for generalizing 
the models to include corrections or to estimate the energy scale for 
such perturbations.

Our multi-scale numerical framework can 
be easily extended to other van der Waals heterostructures 
such as a few-layer graphene stacks~\cite{biGbiG_th1,Moire_Chern}, 
and transition metal dichalcogenides stacks~\cite{Moire_topological,Moire_hubbard}. The effect of mechanical relaxation can also be included in modeling the band structure of twisted trilayer graphene stacks~\cite{Zhu_trilayer_relax}.
Following the scheme of Fig. \ref{TwBLG_multiscale}, 
one can derive models at various levels of approximation 
and scaling of the parameters in conjunction with the relevant symmetry analysis. 
Further experimental probes to the electronic and mechanical properties can also be used  
to refine the numerical models.  This will close the design loop in predicting the 
properties of van der Waals heterostructures,  
towards establishing these systems as a new platform 
for exploring the physics of correlated electrons.

\begin{acknowledgements}
We thank Yuan Cao, Alex Kruchkov, Jong Yeon Lee, Hoi Chun Po, Grigory Tarnopolsky, Yao Wang, Pablo Jarillo-Herrero, and Ashvin Vishwanath for fruitful discussions. This work was supported by the STC Center for Integrated Quantum Materials, NSF Grant No. DMR-1231319 and by ARO MURI Award W911NF-14-0247. The computations in this paper were run on the Odyssey cluster supported by the FAS Division of Science, Research Computing Group at Harvard University.
\end{acknowledgements}

\bigskip

\begin{appendices}

\section{Density Functional Theory Calculations}

The DFT calculations in this work were carried out using the Vienna Ab initio Simulation Package (VASP)~\cite{vasp1,vasp2} with Projector Augmented-Wave (PAW) type of pseudo-potentials, parametrized by Perdew, Burke and Ernzerhof (PBE)~\cite{pbe}. A slab geometry with a 20 \AA $ $ vacuum region is used to reduce the interactions between periodic images. The DFT calculations  are converged with plane-wave energy cutoff $500$ eV and a reciprocal space Monkhorst-Pack grid sampling of size $17 \times 17 \times 1$.

For the mechanical relaxation, the Generalized Stacking Fault Energy (GSFE) was calculated by performing rigid shift of the unit cell and calculate the ground state energy from DFT, and we used a $9 \times 9$ grid in a unit cell for the computation. We fix the in-plane positions and allow the out-of-plane positions to relax. We implemented the van der Waals force through the vdW-DFT method using the SCAN+rVV10 functional~\cite{klimevs2011van, peng2016versatile}. 

The extended Bloch wavefunction basis can be transformed into the maximally-localized Wannier functions (MLWF) basis as implemented in the Wannier90 code~\cite{mlwf}. With this transformation, the effective tight-binding Hamiltonian for a designated group of bands of the material can be constructed. This not only gives an efficient numerical method to reproduce DFT results but also provides a physically transparent picture of localized atomic orbitals and their hybridizations. Our work is based on the systematic analysis of such tight-binding Hamiltonians, which  inherit the {\it ab initio} information without fitting procedures for the numerical parameters~\cite{Bilayer_TBH}. Further corrections for band gaps from advanced GW calculations or other choices of exchange correlation functionals are also compatible with Wannier constructions.


\end{appendices}



\bibliography{TwBLG_bibref2}

\end{document}